\documentclass[]{aastex631}
\usepackage{amsmath,esvect}

\newcommand{\leftshow}[2]{
 \includegraphics[width=#1]{#2}
}

\renewcommand\div{{\bf \nabla} \cdot }

\newcommand\cross{\times}

\newcommand{\field}[1]{\mathbb{#1}}
\newcommand{\C}{\field{C}}
\newcommand{\R}{\field{R}}

\newcommand\rmd{ {\mathrm d} }

\newcommand\deriv[2]{ \frac{\rmd #1}{\rmd #2} }

\newcommand\diff[1]{\,\rmd #1}

\newcommand\vol{{\cal V}}

\newcounter{problemnumber}

\newcommand\eref[1]{equation~(\ref{#1})}

\newcommand\Fref[1]{Figure \ref{#1}}

\begin{document}

\title{Deriving the Topological Properties of the Magnetic Field of Coronal Mass Ejections from In Situ Measurements: Techniques}


\author[0000-0002-0973-2027]{Nada Al-Haddad}
\affiliation{Space Science Center, University of New Hampshire, USA}

\author[0000-0001-7633-3774]{Mitchell Berger}
\affiliation{University of Exeter, UK}

\begin{abstract}
Coronal mass ejections (CMEs) are magnetized plasma systems with highly complex magnetic topology and evolution. 
Methods developed to assess their magnetic configuration have primarily focused on reconstructing three-dimensional representations from one-dimensional time series measurements taken {\it in situ} using techniques based on the "highly twisted magnetic flux rope" approximations. However, the magnetic fields of CMEs is know to have more complicated geometries. Their structure can be quantified using measures of field line topology, which have been primarily used for solar physics research. In this work, we introduce a novel technique of directly quantifying the various form of magnetic helicity within  a CME in the interplanetary space using synthetic {\it in situ} measurements. We use a relatively simple three-dimensional simulation of a CME initiated with a highly-twisted flux rope. We find that a significant portion of the magnetic helicity near 1~au is contained in writhe and mutual helicity rather than just in twist. We discuss the implications of this finding for fitting and reconstruction techniques.
\end{abstract}

\section{Introduction} \label{sec:intro}

Coronal mass ejections (CMEs) are massive intermittent outbursts of magnetized solar plasma. They take place when magnetic field loops or arcades anchored at the solar surface go out of equilibrium and cause magnetic field breakage, with consequently a release of magnetic energy and magnetized plasma material. The exact mechanisms of this loss of equilibrium have been the topic of several studies concerning the complexity of the magnetic field \citep[see, for example, reviews by][]{Forbes:2006,Chen:2011, Gibson:2018,Green:2018}. While limited by observational uncertainties, CME mass and magnetic flux at the time of eruption are typically estimated to be $\sim 10^{16}$~g and $\sim 10^{23}$~Mx, respectively \citep[e.g., see][]{Vourlidas:2000}.

Significant research on the magnetic nature of CMEs has focused on the question of the pre-existence of the magnetic flux rope \citep[e.g., see][]{Gibson:2006}. Measurements from the Solar Dynamics Observatory/Atmospheric Imaging Assembly \citep[SDO/AIA,][]{Lemen:2012} have revealed that hot plasma that appear to correspond to the flux rope can be formed before the eruption \citep[]{Patsourakos:2013,Cheng:2013}. The critical questions then become a) what causes the loss of equilibrium and b) how does the magnetic topology of the erupting CME change during the various phases of the eruption and during the propagation. Other questions remain of high importance in solar and space physics, including understanding whether or not all CMEs follow the same steps of evolution and pattern of behavior. For example, it is not clear whether a hot flux rope forms before every eruption. In this work, we focus on developing a technique that can help to answer the second question, by providing estimates of the various forms of magnetic helicity within a CME in the interplanetary space.

Typically, CME eruption and evolution is thought of as occurring over several phases, from slow evolution and build-up, followed by loss of equilibrium, potentially associated with the onset of specific plasma instabilities \citep[see, e.g.][and references therein]{Demoulin:2010, Temmer:2023}. This loss of equilibrium results in a rapid acceleration of the CME. While a slower acceleration may sometimes last for several hours \citep[]{Zhuang:2022}, in general CMEs are thought to reach their final speed in the mid-corona. At this time, their evolution is determined by propagation \citep[]{Manchester:2017} and interaction with other structures \citep[]{Lugaz:2017}. As such, the loss of equilibrium represents the key transition from the pre-eruptive evolution into the eruption of magnetized plasma. 
Loss of equilibrium can happen  
when the rotation of the footpoints of the magnetic structure cause the magnetic field lines to twist and finally break \citep[]{Fan:2005, Roussev:2007}.  Shearing of the foot points can also cause this breakage, similar to the breakout model of \citet{Antiochos:1999} as further developed in \citet{Lynch:2008}. Flux emergence in the vicinity of the active region may also cause this loss of equilibrium \citep[]{Dacie:2018}. 
The initial magnetic configuration of the filament or flux rope as well as the magnetic and plasma local environment (active region magnetic configuration, for example) determine the magnitude and duration of the motions needed to cause such breakage. Then, the efficiency  of energy transformation, from magnetic, to kinetic and thermal, determines the possibility of an eruption and the final energy of the CME. A review on the influence of different pre-eruptive magnetic configurations on the eruption can be found in \citet{Patsourakos:2020}.
 
The different phases of the solar cycle correspond to different organizations of the global solar magnetic field, with a simpler almost dipolar configuration near solar minimum, with streamers and pseudo-streamers being the most common topological features. Near solar maximum, in contrast, there can be a high number of sunspots, which result in a significantly more complex global solar magnetic configuration. The frequency of CMEs has a strong solar cycle dependence attesting to their magnetic nature \citep[]{Yashiro:2004}. Eruptive CMEs vary significantly in their plasma properties and magnetic field structure both from event to event and over a solar cycle. Their speeds range from about 200~km\,s$^{-1}$ to over 3000 km\,s$^{-1}$ in the corona. There is not one single characteristic that is in common between all CMEs except that they are CMEs. However, their magnetic nature has been always the most significant identifying feature. When measured {\it in situ}, almost all CMEs are characterized by a period of low proton $\beta$ (the ratio of thermal to magnetic pressures) as well lower than expected temperature \citep[for the definition of expecetd temperature, see][]{Lopez:1987}. This resulted in CMEs being referred initially as cold magnetic enhancements \citep[see][]{Burlaga:1979}.


Because they are magnetically dominated, it is critical to understand the magnetic structure and topology of the ejecta portion of CMEs (which should be distinguished from the thermally dominated dense sheath region of the CME). This has been an intense area of research for the past forty years, particularly since the realization that magnetic ejecta are large-scale coherent structures \citep[]{Burlaga:1981}. Models of the magnetic field inside CMEs include force-free \citep[]{Lepping:1990} and non-force-free \citep[see for example][]{Nieves:2016,Nieves:2018b} cylindrical models, models with a toroidal cross-section \citep[e.g., see][]{Marubashi:1986,Romashets:2003}, as well as models based on the magneto-hydrostatic approximation \citep[relying on the Grad-Shafranov equation][]{Hu:2002,Moestl:2008}, also with a toroidal approximation \citep[]{Hu:2017}. Magnetic fields inside CMEs are primarily measured before the eruption through photospheric magnetographs that provide line-of-sight or vector magnetograms close to the solar surface; and after the eruption through magnetometers onboard satellites outside of Earth's magnetosphere. Because there are less than 10 spacecraft with magnetometers within a 4 $\pi$ sphere of 1~au radius around the Sun, most CMEs are only measured {\it in situ} by one, rarely two satellites \citep[see, for example][]{Kilpua:2011, Lugaz:2018, Regnault:2023b, Lugaz:2024}, with separations large enough to be meaningful to understand the CME global nature. As such, fitting and reconstruction models are necessary to extract global properties from these local measurements. However, previous research has shown that such techniques come with limitations associated with their assumptions \citep[]{AlHaddad:2011,AlHaddad:2019} and, in addition, different models are rarely found to be consistent with each other \citep[]{Riley:2001,AlHaddad:2013}. The principal limitation encountered in these models pertains to their partial consideration of the magnetic field's overall behavior, predominantly focusing on its twist component while neglecting all other components of the magnetic helicity.

Magnetic helicity is a fundamental quantity for quantifying the magnetic field of CMEs as it is quasi-conserved in a closed system, even in the presence of reconnection, making it a more adequate scalar quantity to describe magnetic fields than magnetic energy. 
It is a measure of the topology, linkage, and entanglement of multiple magnetic field lines in a system. Such as, it represents various features of magnetic fields: twisting, writhing, coiling, knotting and linking among others.  
To be properly quantified, magnetic helicity requires a knowledge of the 3-D structure of the magnetic field, or alternatively, a time history of the helicity flux into an initially untwisted field (e.g. a potential field). The 3-D structure could be provided, for example by vector magnetogram measurements of the solar surface, as well some further assumptions about the magnetic field, such as being close to force-free. 


In general, helicity, $H$,  is defined as a triple integral, which requires knowledge of the magnetic field throughout a closed volume. For a closed volume ($V$),
\begin{equation}
H = \int _V {\bf A} \cdot {\bf B} \, dV
\end{equation}
where the magnetic field can be written ${\bf B} = \nabla \times {\bf A}$ in terms of the vector potential (${\bf A}$). In a system, two kinds of helicities can be identified: self-helicity and mutual helicity. Mutual helicity occurs when the projection of two field lines in a plane intersect (for example sheared arcades above a low-lying filament) and is fundamentally 3-D and requires multiple field lines. Self-helicity can take the form of twist or writhe. Twist represents the amount of turns that a magnetic field line does around an axis, whereas writhe represents how this axis itself may contain helicity. In the absence of writhe, twist can be thought of as a pseudo-2-D form of helicity as the axis introduces an invariance in the system. How magnetic helicity goes from one form to another is a fundamental problem in plasma and solar-terrestrial physics. For example, the well-studied kink instability is a transformation of helicity from twist to writhe \citep{Torok:2003}. However, although all these forms of helicity are well documented, nearly all CME fitting and reconstruction techniques focus exclusively on twist.

Solar research often focuses on helicity changes, which require estimations of the velocity on the boundaries to be computed. Multiple techniques have been designed to calculate helicity from solar magnetic field measurements \citep[]{Berger:1984,Nindos:2003, Demoulin:2003,Kusano:2004,Pariat:2017}. Such techniques however, have not been used or developed for helioshperic research where 2-D measurements are  not available. In this work, we develop a technique that quantifies the magnetic helicity inside a CME in interplanetary space, we apply it to a numerical simulation, where the 3-D information of the magnetic field is accessible.

The rest of the article is organized as follows. In Section~\ref{sec:insitu}, we describe the simulation used in this study. In Section~\ref{sec:separatrix}, we describe the method used to detect separatrices within the simulation. In Section~\ref{sec:helicity}, we describe how we will quantify aspects of topological structure within the simulation. These aspects include winding numbers, twist and writhe, self helicities of individual flux regions and mutual helicities between flux regions, and total helicity. In Section~\ref{sec:application}, we apply the techniques discussed to the simulation. We conclude and discuss the consequence of our methods in Section~\ref{sec:conclusion}.




\section{CME Simulation Used to Quantify Helicity in Interplanetary Space}\label{sec:insitu}

{\it In situ} measurements from spacecraft at the L1 point, in planetary orbits (such as MESSENGER and Venus Express) or in heliocentric orbits (such as STEREO, Solar Orbiter and Parker Solar Probe) measure the three components of the magnetic field as a time series. Under the assumption that temporal changes occur on a slower timescale than the transit of the structure over the spacecraft, these measurements provide information on a one-dimensional curve inside the CME volume. The validity of this assumption near 1~au has been recently further investigated \citep[]{Regnault:2023b,Regnault:2024b} but this is beyond the scope of this work. 
This information may be three-dimensional (e.g. the three components of the magnetic field), but does not directly tell us about the magnetic field away from the spacecraft trajectory (although electric current measurements help with derivatives of the field but only on local scales). The common procedure is to assume some shape for the CME, for example a single twisted flux tube \citep{Hu:2002,Hu:2022,Jin:2017,Torok:2022}, and find the best fit for flux tube parameters (e.g. net flux, twist, etc.). There are several flux tube models to choose from (e.g. uniformly twisted, linear force free, etc.).

However, magnetic clouds may have a more complicated structure than a single simple flux tube \citep{Farrugia:2011,Ledvina:2023}. Magnetic fields ejected from the solar corona may consist of two or more braided tubes for example, or a central flux tube with an overlying arcade. Also, magnetic reconnection with the ambient interplanetary magnetic field (IMF) may complicated the magnetic field inside the magnetic cloud further, either through magnetic erosion \citep[]{Dasso:2006} or by adding magnetic flux onto the erupting CME. Some advanced simulations \citep[]{Torok:2018,Regnault:2023a} can produce relatively complex magnetic field structures for the CMEs, but this brings up the following question: how do we compare the magnetic structure found in different simulations, both to each other and to observations, in a quantitative manner? In addition, how do we quantify the change of the magnetic helicity in a simulation during the CME propagation?


\subsection{Simulation} \label{subsec:simulation}
Our approach in this work is to use a 3-D numerical simulation of a CME from which we can extract all required magnetic field data to perform our technique. Here, we use the simulation of a CME which is propagated from the Sun to Earth as recently published in \citet{Regnault:2023a}. We give a quick summary of the simulation set-up and models used below, but interested readers can refer to \citet{Regnault:2023a} for more information.

\begin{figure}[hbt!]
\begin{centering}
\includegraphics[width=13cm]{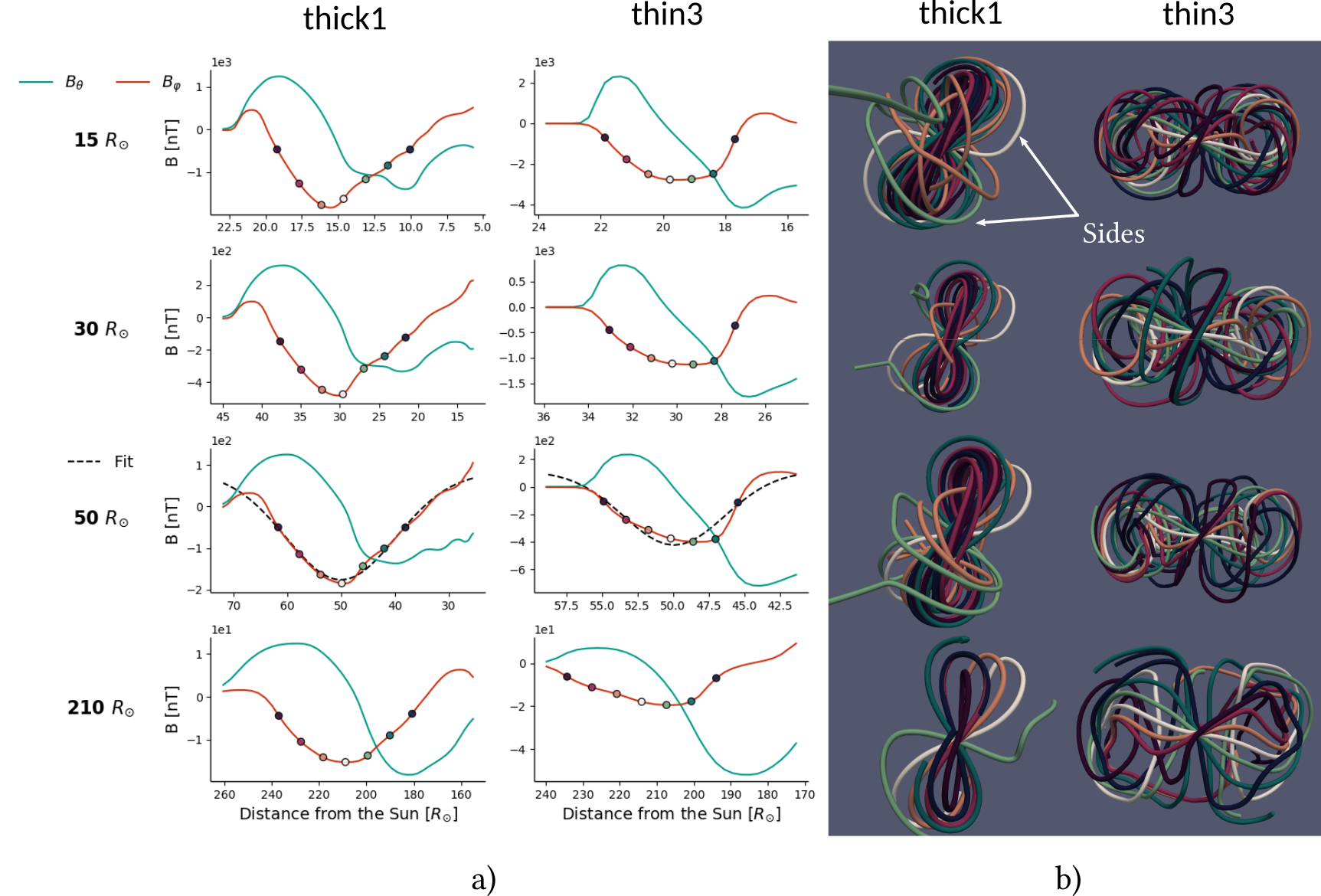}
\caption{A visualization of the magnetic field structure inside the simulated CME as well as 1-D cuts through the magnetic field at different distances. Adapted from \citet{Regnault:2023a}. }
\label{fig:simu}
\end{centering}
\end{figure}

The PLUTO code is used to solve the 3-D MHD equations in an adaptive mesh refinement (AMR) spherical grid from the low corona (1~$R_\odot$) up to 2~au. The AMR grid allows to keep a relatively high spatial resolution exclusively around the propagating flux rope. The background solar wind and interplanetary magnetic field in the simulation are initialized with a polytropic Parker solar wind solution along with a dipole for the magnetic field configuration of the Sun (similar to that of the quiet phase of the solar cycle) and uses an ideal equation of state. To heat the solar wind, the ratio of specific heat $\gamma$ is set to 1.05 which produces an almost isothermal corona at about $10^6$~K. The solar wind model is calibrated such that the simulated solar wind speed, mass loss rate and angular momentum match the ones deduced from observations at 1~au.

To model the magnetic structure of the CME, we use the modified Titov-D\'emoulin flux rope model \citet{Titov:2014} based on the flux rope model of \citet{Titov:1999}. This allows an analytical formulation of a flux rope magnetic structure which can be inserted onto the solar wind plasma and magnetic field. The flux rope is initially out of equilibrium and thus erupts quickly after its insertion. The initial magnetic field of the flux rope is 22~G. The flux rope structure is positioned on the equator of the Sun aligned with the solar equatorial plane. The simulation is then evolved in time from this initial set-up. A 3-D view of the magnetic field lines inside the CME as it reaches 1~au is shown in~Figure~\ref{fig:simu}.
The simulation used in this work is the {\it thick-1} simulation from \citet{Regnault:2023a}.

\section{Separatrix structure}\label{sec:separatrix}

In this section, we provide some mathematical methods and tools for measuring the field structure within a volume. We divide the field into sub-regions separated by separatrices (or quasi-separatrix layers) \citep{Berger:1989,Priest:1995,Tassev:2017}. Within each sub-region, we can calculate the net magnetic flux. The magnetic field lines  in most cases cross the boundary, so we employ open helicity techniques to define the magnetic helicity. We  do this by summing winding numbers between individual field lines, weighted by the magnetic flux. This gives the self-helicity of each sub-region. Moreover, we can divide the self-helicities into twist and writhe contributions. If there is more than one region, we can then obtain mutual helicities between the regions. Summing self and mutual helicities gives the total helicity of the volume.

Two neighboring magnetic field lines usually stay close to each other. However, in the presence of either \emph{current sheets} or \emph{magnetic null points}, neighboring lines can quickly diverge. We divide the magnetic field domain into sub-regions where any two neighboring field lines remain close. The boundaries of these regions are called \emph{separatrices}.

These ideas align closely with the idea of \emph{field line mapping}, see for example \citet{Titov:2007}. Suppose for simplicity a magnetic field stretches between two parallel planes at $z=0$ and $z=L$. Then there is a mapping
of points on the bottom plane to the top plane (starting point of a field line to its endpoint)
\begin{equation}
 f:\{z=0\} \rightarrow \{z=L \}, \hspace{1cm} (x_0, y_0) \rightarrow (x_L, y_L).   
\end{equation}
Discontinuities in this mapping arise along the separatrices. 

In most situations, there will be strong thin layers of current (\emph{current layers}) rather than zero thickness current sheets,
so the field line mapping will have steep gradients rather than be discontinuous. In these situations one often refers to \emph{quasi-separatrix layers}, or \emph{QSLs} \citep{Berger:1989,Priest:1995,Tassev:2017}. 

To detect the separatrices or QSLs, we look at the field line mapping. Suppose the field resides in a volume $\vol$, with boundary $S$. Create a grid of points on $S$, and calculate the field lines which emanate from points on $S$ where the magnetic vector points inwards. Let the distance between neighbouring starting points be $\delta_0$. Now field lines can diverge from each other provided that the field strength changes to conserve flux and keep $\div \vec B = 0$ (e.g. an expanding field will become weaker). One expects, for uniform expansion in $x$ and $y$,
\begin{equation}
\delta_0^2 |\vec B_0| \approx \delta_L^2 |\vec B_L|. 
\end{equation}
We can then detect separatrices or QSLs by finding a ratio  $\delta_L\delta_1$ which deviates strongly from this rule.

Thus, we test all pairs of neighbours using Graph theory algorithms. Each field line will be a vertex in the graph, and an edge is added for each pair of field lines which remain close to each other. One then looks for subsets of the graph which are connected -- these correspond to the field line regions. Fast linear time algorithms exist to divide a graph into its connected subsets (e.g. \emph{breadth-first or depth-first searches}) \citep{Hopcroft73}.




\section{Quantifying the magnetic field topology} \label{sec:helicity}
The \emph{magnetic field topology} within a volume $\vol$ refers to features of the field line geometry which do not change if the field evolves due solely to ideal motions which vanish on the boundary of $\vol$. In general, a magnetic cloud will have open lines (lines which cross the boundary of the cloud), so we will concentrate on the winding structure. 

For two individual field lines with endpoints on the boundary, we will define their \emph{winding number} below. If we have a set of field lines, we can sum the winding numbers between all pairs of field lines. If we weight this sum by flux, then we obtain the \emph{self helicity} of that set. Sometimes it is useful to forego the flux weighting, to obtain the magnetic winding \citep{Mactaggart:2021,Prior:2021}. 
The self helicity for a flux concentration in the shape of a flux tube can be decomposed into \emph{twist} of the field about the central axis of the tube, and the \emph{writhe} of the central axis.
For two magnetic regions we can sum the windings between pairs of field lines, one in each region and weighted by flux, to obtain their
\emph{mutual helicity}. 

We describe the \emph{magnetic helicity} as the net winding or linking of field lines within the volume $\vol$ \citep{Moffatt69, Berger:1984, Pevtsov2014, prioryeates14}. If the field consists of several distinct flux regions, say separated by separatrices, then the magnetic helicity will be the sum of the self and mutual helicities.

\subsection{Winding of two curves stretching between two planes}
The simplest and most intuitive example of winding occurs when two curves travel between parallel planes.
Suppose a single curve travels upwards from the plane $z = 0$ to the plane $z = L$. Then we can ask how much it winds about the central axis $z = 0$
Let $(r, \theta, z)$ be cylindrical coordinates. Then $\hat \theta = \hat z \cross \hat r$, so summing over all changes in $\theta$ from the bottom to the top gives a winding number
\begin{equation}
W = \frac 1 {2\pi} \int_0^L{\frac{1}{r^2}\hat z \cross \vec r \cdot {\deriv {\vec r} z}\diff z}.
\label{eq:wz}
\end{equation}
Here one complete right handed  turn gives $W=+1$.

\begin{figure}
\begin{centering}
\leftshow{18 cm}{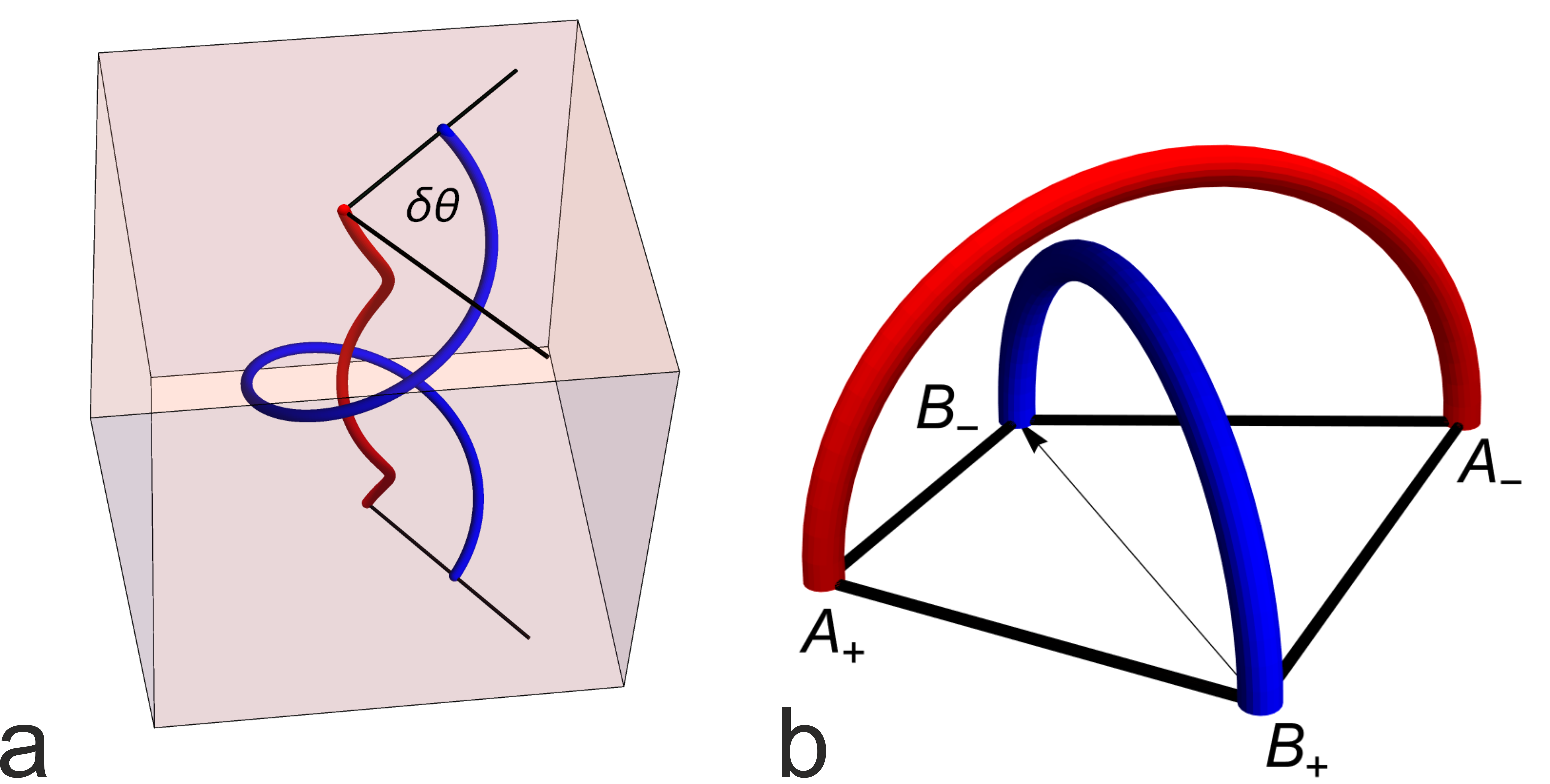}
\caption{
a) Winding number of two curves 
about each other. Here the 
winding is  $W = \delta \theta +\cal{T}$,
where ${\cal T}=1$ gives the number of extra turns.
b) Two arches above a plane. The sum of the angles at $A_+$ and $A_-$ (or equivalently $B_+$ and $B_-$) determine the winding number.}
\label{fig:relativewinding}
\end{centering}
\end{figure}

Next suppose two curves travel upwards from the plane $z = 0$ to the plane $z = L$. We can then ask how much they wind about each other. In particular,
how much does the relative position vector $
\vec r_{12}(z) = \vec r_2(z) - \vec r_1(z)$
rotate as we go from the bottom to the top?
(see \Fref{fig:relativewinding}a). Here we replace $\diff \vec r /\diff z$ with  $\diff \vec r_{12} /\diff z$ in \eref{eq:wz} to capture the rotation of this relative position vector. Thus
\begin{equation}
W = \frac 1 {2\pi} \int_0^L{\frac{1}{r_{12}^2}\hat z \cdot \vec r_{12} \cross {\deriv {\vec r_{12}} z}\diff z}.
\label{eq:wzrel}
\end{equation}
We can also obtain $W$ from knowing the angle of $\vec r_{12}$ with respect to the $x$ axis at both the top and bottom curves. The difference between the angles at top and bottom (plus some integer ${\cal T}$ to account for complete windings) gives $W$ (see \Fref{fig:relativewinding}a).

What if some field lines do not always go upwards, i.e. they have local maxima and minima in $z$? In this case we can cut the field line into pieces beginning or ending either at $z=0$, $z=L$, or at each maxima or minima. We then add up the contributions of each piece to the total winding number (see
\cite{BergerPrior2006} for details). 

In either case (field lines extending between two planes or looping back to one plane) the winding number is a function of the four end point positions, plus an integer $\cal T$ measuring any extra turns of one line about the other.

\subsection{Winding numbers for two arches above a plane}

Suppose that all four endpoints reside on the same plane.
Suppose curve $A$ has endpoints at positions $A_+$ and $A_-$,
while curve $B$ has endpoints $B_+$ and $B_-$ (see Figure~\ref{fig:relativewinding}b). Imagine that initially curve $B$ emerges into $\vol$ (like a rising dipole) at position $B_+$. Then the negative endpoint  moves to $B_-$. As seen from $A_+$, it sweeps through an angle
$\angle B_+A_+B_-$. Similarly, $A_-$ sees $\angle B_+A_-B_-$. The net winding will then be \citep{Berger:2009}
\begin{equation}\label{windformula}
    W = \frac{1} {2\pi} \left(\angle B_+A_+B_- - \angle B_+A_-B_- \right).
\end{equation} 
Note that an angular motion is positive if it moves anti-clockwise about a vertical normal. 
It will be useful in our analysis to sometimes measure angles against some some common reference direction (e.g. the $x$ axis), so that rather than looking at two angles, we are looking at the orientations of the four sides of the quadrilateral $A_+ B_+A_-B_-$:
\begin{equation}
\label{windingformula}
    W = \frac{1} {2\pi} \left(\angle \vv{A_+B}_- - \angle\vv{A_+B}_+\right) - \left(\angle \vv{A_-B}_-
    -\angle \vv{A_-B}_+ \right).
\end{equation} 

\subsection{Winding numbers in a Cartesian or spherical region}

Next suppose that $\vol$ is some finite sub-volume of space. In this paper, we will give an example where $\vol$ consists of a region in spherical coordinates $r_0<r<r_1$, $\theta_0<\theta<\theta_1$, $\phi_0<\phi<\phi_1$. 

We suggest that the definition of winding number for any two field lines with endpoints on the boundary should obey some simple rules. Firstly, it must be invariant to deformations of the curves where the deformations vanish on the boundaries. Secondly, it should be a function of the four endpoints, plus an integer. Thirdly, it should change continuously if the endpoints move on the boundaries. Fourthly, it should change sign if we reverse the direction of one of the field lines.

We also suggest that the definition should have a simple geometric or physical meaning. For example, relative magnetic helicity measures how much the topological complexity within a volume arises from electric currents within the volume, as it measures helicity relative to a potential field. We will not use relative helicity here, however, as it would be computationally complex to calculate for each pair of field lines.

Instead, we appeal to the idea of winding as measuring the angle through which one curve rotates about the other. 
In general, the helicity of two flux tubes arises from the internal twist of the tubes, plus the winding between the tubes, and more often than not, helicity enters a volume when the endpoints spin about their axis, or move about other endpoints. Thus the helicity flux consists of spin terms and orbit terms. The orbit terms between two flux tubes can be used to determine how the winding number between their two axis curves evolve.

To calculate winding, we can first consider the helicity of two untwisted tubes which wrap about each other. 
We will first assume that the axes of the tubes $A$ and $B$ are simple straight line segments (or arches if both endpoints are on the same plane).
Later we can add an integer $\cal T$ if the true curves have any full turns in addition to the winding angle between the simple line segments.

Consider the first term in equation (\ref{windformula}). Imagine again that both footpoints of the $B$ tube emerge together as a dipole at the position of $B_+$, and then the negative footpoint moves to $B_-$. The existence of the footpoint at $A_+$ generates the orbit term represented by the angular change during the motion. To obtain the helicity flux, we integrate $\vec A \cdot \rm d \ell$ along the path of the moving footpoint. Once the helicity flux has been found, we can divide by the magnetic flux of $A_+$ to find the winding number.

We employ the \emph{absolute helicity} method described in \cite{Berger:2018}. Here, the vector potential $\vec A$ is taken to be the vector potential for the poloidal field of $A_+$. For volumes bounded by a single plane or sphere, or bounded  on two sides by parallel planes or concentric spheres, the relative helicity will match the absolute helicity. In less symmetric volumes there will be some differences. 

\subsubsection{Spherical Boundary}   

\begin{figure}[tb]
\begin{centering}
\includegraphics[width=5.2cm]{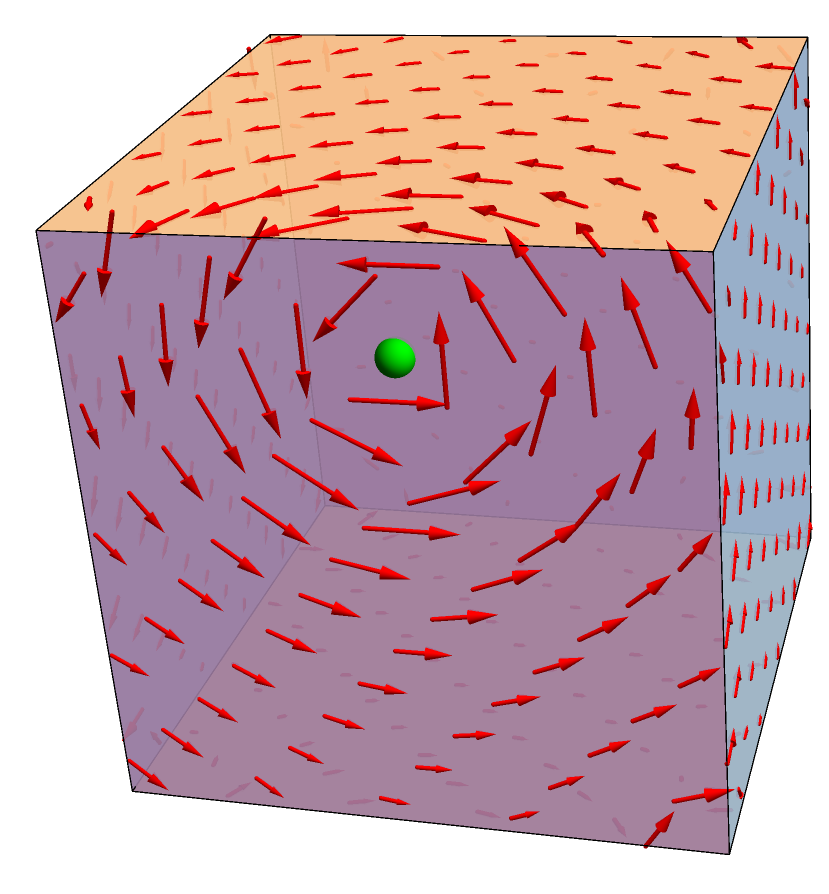}\hskip 0.5cm
\includegraphics[width=5.2cm]{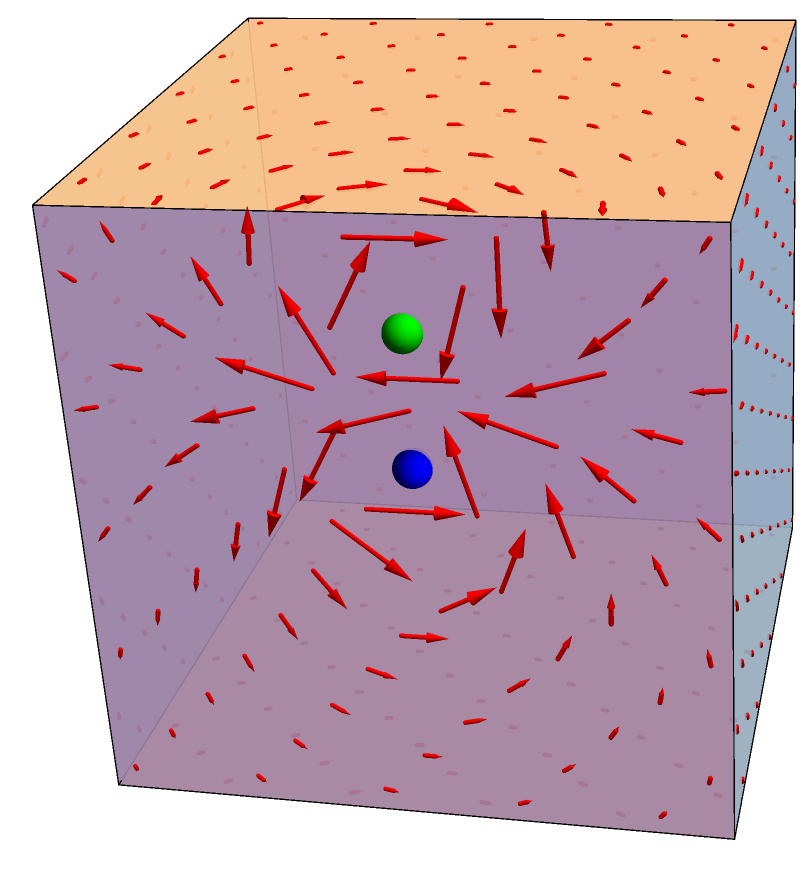}
\includegraphics[width=6.5cm]{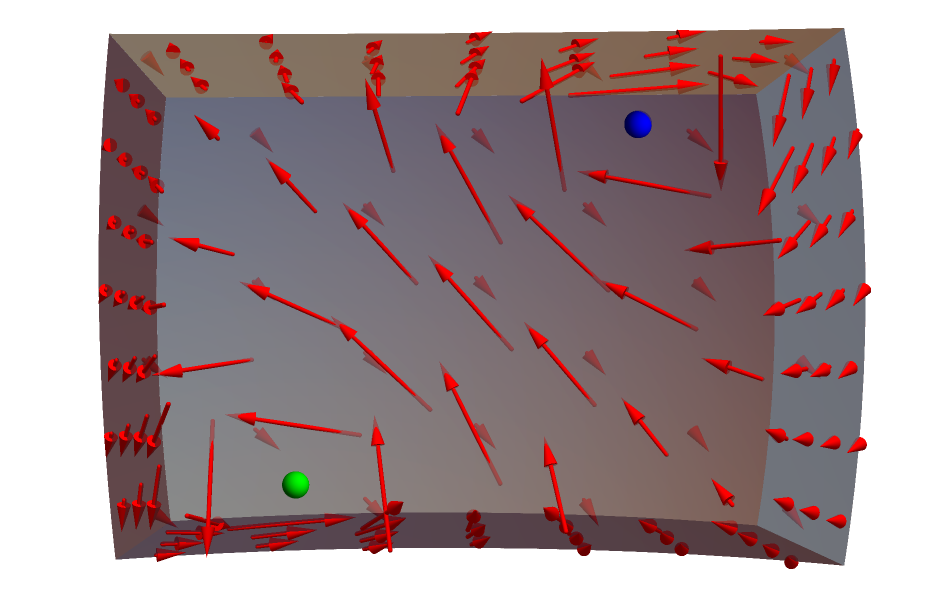}
\caption{The left panel shows the vector potential on a cube for a single footpoint, with return flux at the corners. The middle panel shows the vector potential for a dipole. The right panel shows a dipole example on a spherical region.}
\label{fig:afield}
\end{centering}
\end{figure}

We will need to modify our orbit terms in the winding number equation. There are a few equivalent methods of doing this. One is to find the appropriate  vector potentials which give the absolute helicity flux when the footpoints of tube $B$ move into position about tube $A$. For the angular term involving $A_+$, we can ignore $A_-$ for the moment, and distribute a return flux around the surface proportional to the curvature at each point \citep{Campbell:2014,Xiao:2023}.
We can then do the same for $A_-$ with the opposite return flux. This enables us to calculate the two terms in eqn. (\ref{windformula}) independently. For a sphere, the vector potential for 
$A_+$ at the North Pole is
\begin{equation}
   \left( \frac{1+\cos\theta}{2}\right)\frac{\Phi_A}{2\pi r \sin \theta}\hat{\phi}. 
\end{equation}
The $(1+\cos\theta)/2$ factor represents the return flux above co-latitude $\theta$.

\subsubsection{Cubic Boundary}
Next consider field lines inside a cube. Suppose we have a footpoint on the boundary at $A_+$. Again flux tube $B$ emerges at the position of $B_+$ and $B_-$ moves to its final position. The helicity flux (divided by $\Phi_A$) gives the winding of tube $B$ about $A_+$. We employ the poloidal field method of \cite{Berger:2018}. However, we need a return flux for $\Phi_A$. 

There are two simple methods of providing the return flux which, interestingly, give the same result in our numerical calculations. In the first method, move a footpoint with equal and opposite flux to $A_+$ to every position $\vec x$ on the numerical grid for the cubic boundary. at each position calculate the average $\vec A$ near $\vec x$. The contribution of the second footpoint to this average vanishes, as the vector potential near a footpoint consists of concentric circles centred on the point. In the second method, the return flux is distributed according to curvature, as in the spherical case. For a cube, this means that each corner of the cube receives $-1/8\Phi_A$. We found that the two methods agree, up to a small numerical error proportional to $1/N$ (where each side receives $N^2$ grid points).

Of course, when the second footpoint for tube $A$ is included, the need for return flux vanishes. Nevertheless, it is to understand how individual footpoints contribute to $\vec A$ and hence winding integrals. 
Figure \ref{fig:afield}  shows an example with one and then two footpoints for flux tube $A$.

We calculate the winding numbers between any pair of flux tubes by calculating the boundary vector potential of one of the tubes, and integrating this potential from one footpoint of the second flux tube to the other. This gives the winding number up to an integer. We then compare with the winding of the two curves calculated using the methods of Figure~\ref{fig:relativewinding} and compare. The integer which gives the closest match is then added to the vector potential result.

\subsection{Helicity}
The magnetic helicity of a region or subregion can now be calculated as the sum, weighted by magnetic flux, of winding numbers for all pairs of field lines: let $\vec x_1 = (x_1, y_1)$  and $\vec x_2 = (x_2, y_2)$ be footpoints at $z=0$. Then the helicity is 
\begin{equation}
  H = \int_{z=0} \int_{z=0} W(\vec x_1,\vec x_2) B_z(x_1, y_1) B_z(x_2, y_2) \rmd ^2 x_1 \rmd^2 x_2.\label{helicityintegral}  
\end{equation}
For a volume residing between parallel planes or concentric spheres (or a half space, the interior of a sphere or the exterior of a sphere) the sum of the winding numbers equals the relative helicity \cite{Berger:2018}. For the volumes considered in this paper (a Cartesian cube or a cuboidal region in Spherical coordinates) this may not be true: the helicity of a field which is potential both interior and exterior to the volume may be nonzero, and may not be measured by the winding number technique. However, we we still employ the winding number technique as it provides a good measure of semi-local topological structure within the volume. An alternative technique for measuring topological structure within a subvolume of space, using wavelets, is given in \citep{Prior:2020}.

\subsection{Self and Mutual Helicity}
Suppose the separatrix structure gives us $N$ subregions of the volume. The double integral for the helicity, equation \ref{helicityintegral}, can be divided into subintegrals for each pair of subregions, i.e.
\begin{equation}
 H= \sum_{i=1}^N H_{ii} + \sum_{i=1}^N \sum_{j=1, j\ne i}^N H_{ij} 
\end{equation}
Here in $H_{ij}$,  $\vec x_1$ lies in the $i$ subvolume, and $\vec x_2$ lies in the $j$ subvolume. 

The double integrals over the same subvolume $H_{ii}$ are called the \emph{self helicities} while the integrals linking different subvolumes are called the \emph{mutual helicities}. By expressing the helicity in terms of winding numbers we can readily divide the helicity into self and mutual parts. The self helicities will tell us which regions are most helical, and whether some regions have positive helicity whereas others have negative helicity. The mutual helicities will measure how much different regions wrap around each other (for example, see figure \ref{fig:mh} below).

\section{Application of the Techniques to Quantify the Helicity for the Simulated CME}\label{sec:application}

\subsection{Regions, Self and Mutual Helicities}

\begin{figure}[hb]
\begin{centering}
\includegraphics[width=17 cm]{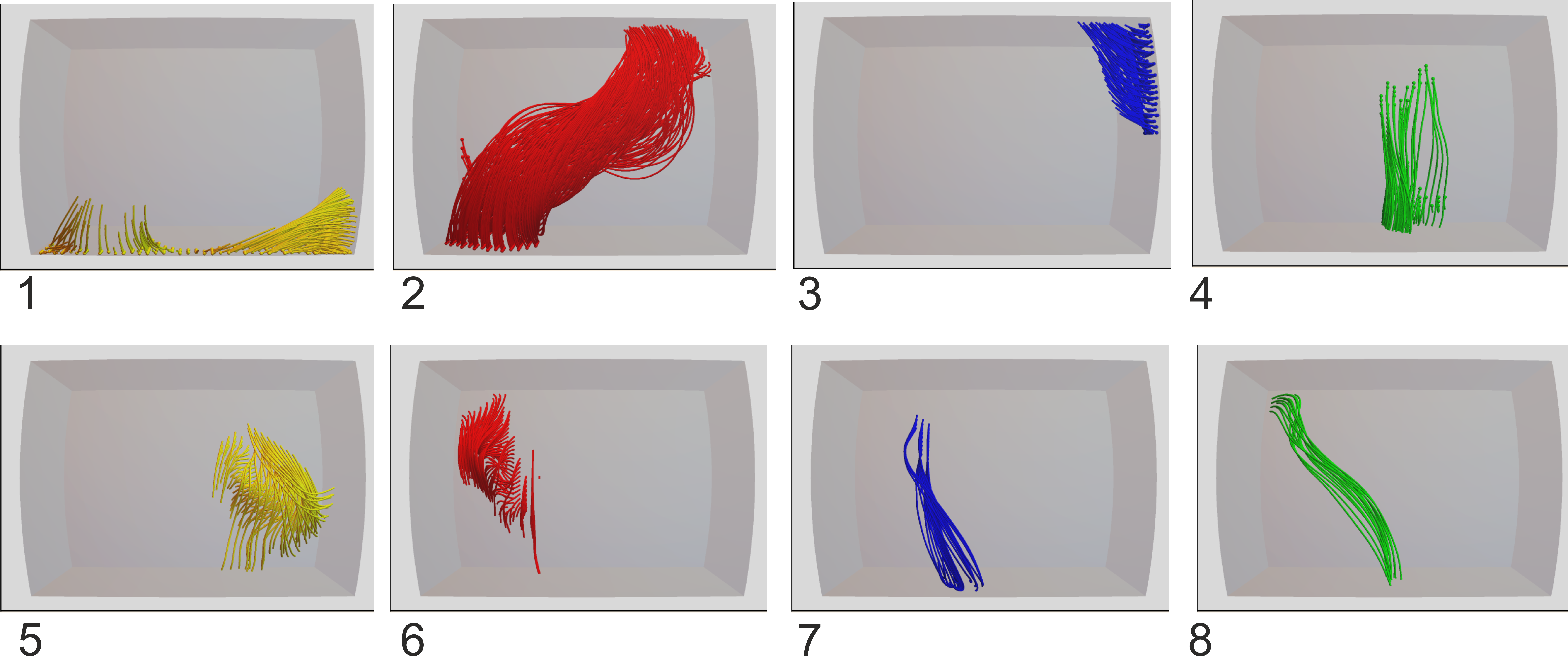}
\caption{Eight groups of magnetic field lines divided by separatrices for the CME simulation of \citet{Regnault:2023a} near 1~au. Five negligible groups are not shown. The horizontal coordinate corresponds to solar longitude $ 2.62< \phi < 3.65$. The vertical coordinate corresponds to solar latitude $1.22<\theta< 1.91$. The radial coordinate spans $180 R_\odot< r< 237R_\odot$. }
\end{centering}
\label{fig:sphericalgroups}
\end{figure}

\begin{figure}[b]
\begin{centering}
\includegraphics[width=8cm]{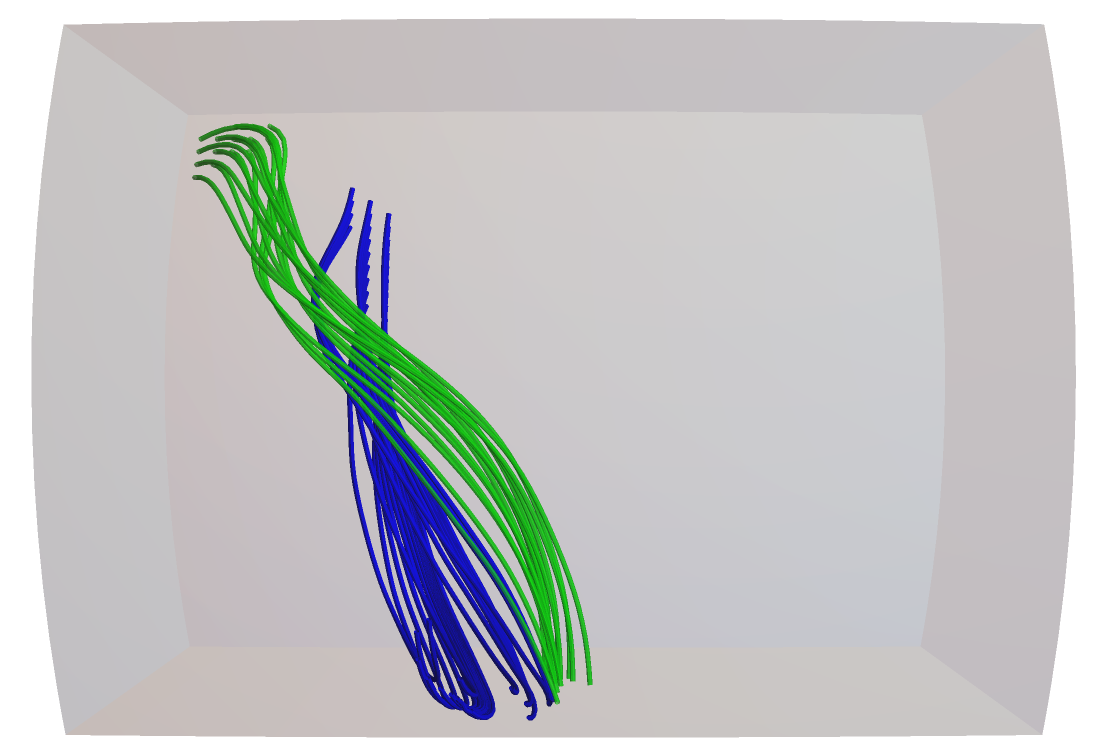}
\caption{Groups 7 (Blue) and 8 (Green) wrap around each other with an average winding number of -0.75, with $H_{78}/{\Phi_7\Phi_8}= -1$. The mutual helicity between these two regions is comparable to their self-helicity.}
\label{fig:mh}
\end{centering}
\end{figure}

Following the techniques described in Section~\ref{sec:separatrix}, we first identify the quasi-separatrix layers to divide the simulation into separate regions of magnetic field lines. We identify 13 separate regions, out of which five have a very weak magnetic flux and small number of field lines. Figure~\ref{fig:sphericalgroups}  shows the eight groups of magnetic field lines from the {\it thick} simulation grouped in the manner described in Section~\ref{sec:separatrix}. The core of the magnetic ejecta is in region 2, which has the highest net magnetic flux (8.6 $\times 10^{-2}$~Mx). While regions 7 and 8 have relatively weak net fluxes ($\sim$  5.5 $\times 10^{-3}$ and 1.8 $\times 10^{-3}$ Mx, respectively), they are important to visualize the influence of mutual helicity as shown in Figure~\ref{fig:mh}. In fact, regions 1, 3, 4, 5, and 6 all have small self-helicity but contribute to the mutual helicity of the volume. Their net magnetic fluxes are 2.9 $\times 10^{-2}$, 1.1  $\times 10^{-2}$, 1.6  $\times 10^{-2}$, 4.1  $\times 10^{-2}$ and 2.7  $\times 10^{-2}$~Mx respectively. 

Then, following the techniques described in the previous section, we derive the self-helicity for each region and mutual helicity for each pair of regions. Table \ref{table:selfmutual} gives the self and mutual helicities (divided by the product of the fluxes) for each group or each pair of groups. It can be seen that regions 7 and 8 have high mutual helicity ($\sim$ -1 $\times$ their fluxes), but it is also the case for region 2 with regions 7 and 8 and to a lesser extent with regions 3 and 4. Overall, taking into consideration the magnetic fluxes, the sum of the self-helicity is only about 71\% of the total helicity, therefore neglecting the mutual helicity introduce an underestimating by about 29\% of the total helicity of the system. The main contributions to the mutual helicity are between region 2 on the on hand and regions 1, 3, 4, 6, 7, 8 on the other hand, but there are also non-negligible contributions of the mutual helicity between regions 1, 2 and 4 on the one hand and region 5. Almost all the self-helicity is contained in region 2, due to its high net magnetic flux. We further discuss the consequence of the repartition of the helicity between self and mutual helicities in Section~\ref{sec:conclusion}

\begin{table}[ht]
\begin{center}
\begin{tabular}{||c c c c c c c c c||} 
 \hline
{\it Group}  & 1 &2 &3 &4 &5 &6 &7 &8 \\ [0.5ex] 
 \hline\hline
 1 &  {\bf -0.02} &-0.18  &-0.01 &0.06 &0.04 &0.02 &0.12 &0.07 \\ 
 \hline
 2 &  & {\bf -0.57} &-0.317 &-0.254 &0.012 &-0.05 &-0.89 &-0.95 \\
 \hline
 3  & & & {\bf 0.08} &0.117 &0.049 &0.048  &0.15  &0.18 \\
 \hline
 4  & & & &{\bf 0.02} &-0.146 &0.045 &-0.11 &-0.02\\
 \hline
 5  & & & & &{\bf -0.01} &-0.004 &-0.12 &-0.04 \\
 \hline
 6  & & & & & & {\bf -0.01} &-0.005 &-0.14 \\
 \hline
 7  & & & & & & & {\bf -0.30} &-0.96  \\
 \hline
 8  & & & & & & &  &{\bf -0.47}\\[0.5ex]
 \hline \end{tabular}
 \caption{ Self (on the diagonal) and mutual helicities between the groups (divided by product of fluxes).}  
\label{table:selfmutual}
\end{center}
\end{table}


\subsection{Twist and Writhe}
For flux regions in the shape of coherent flux tubes, we can further decompose the self helicities into twist and writhe \citep{AlHaddad:2011, AlHaddad:2019}. Techniques for calculating the writhe can be found in \citep{BergerPrior2006} and are quickly summarized below.

\begin{figure}[h]
\begin{centering}
\leftshow{8 cm}{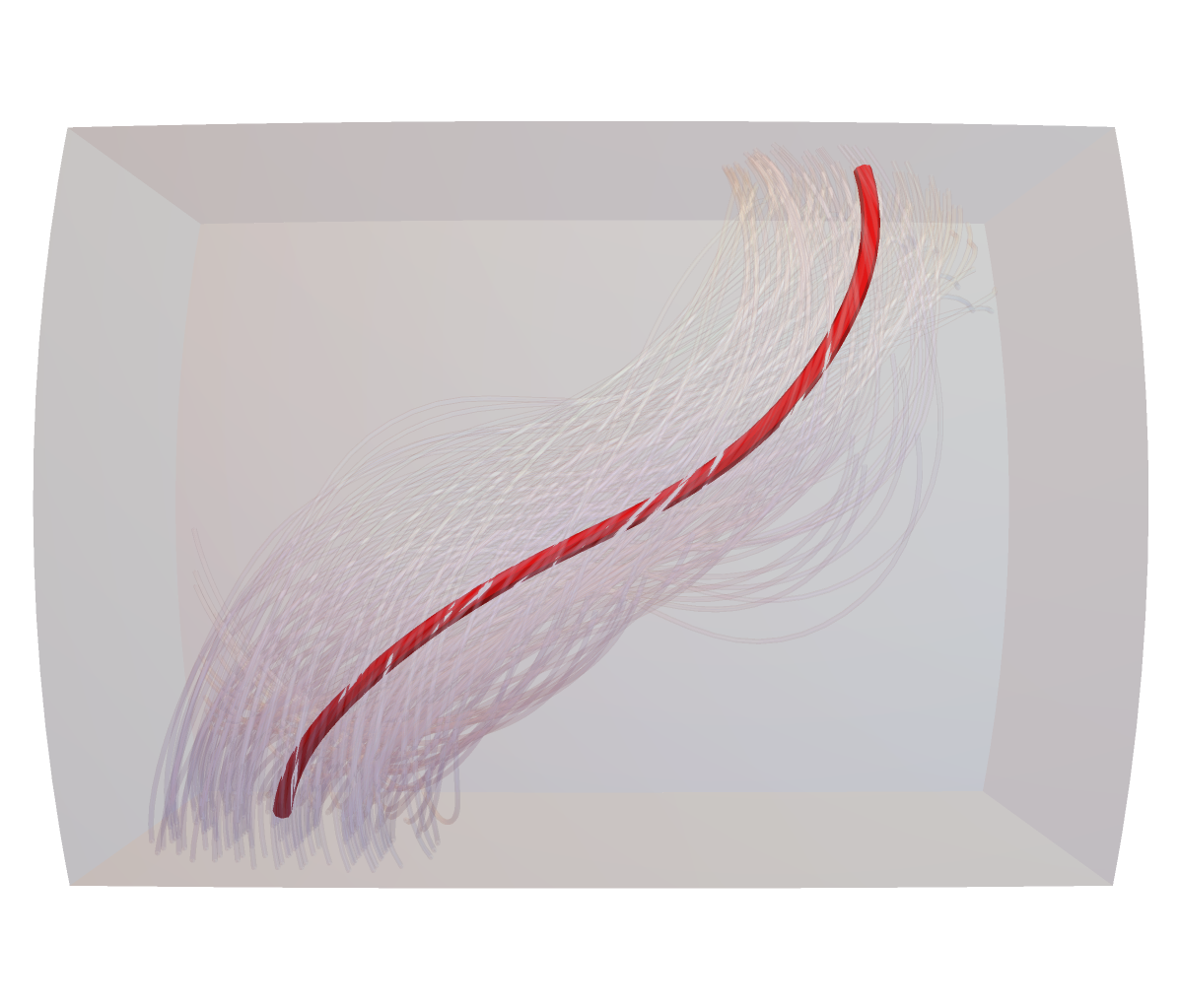}
\caption{
Group 2 magnetic field lines along with its central axis. The axis has writhe, which is further discussed in the text. This group represents the core of the magnetic ejecta as the CME reaches 1~au.
}
\label{fig:tubeaxis}
\end{centering}
\end{figure}

Suppose we project two closed linked curves onto a plane. If, say, the linking number is ${\cal L} = - 2 $, then for any projection (or viewing) angle we will see four negative crossings (or, e.g. 7 negatives and 3 positives). Thus ${\cal L} = 1/2 (C_+-C_-)$, where $C_+$, $C_-$ give the number of positive and negative crossings. For a single writhed curve, the signed number of crossings $C=C_+ -C_-$ will vary with projection angle. Here the writhe equals the average of $C$ over all projection angles (i.e. over a solid angle of $4\pi$) \citep{Klenin:2000}. The winding number between parallel planes is similar -- but here 
$
W = \frac 1 2 \langle C\rangle 
$ 
where we average $\langle C\rangle$ only over the $2\pi$ projection angles parallel to the planes. 

The Calugareanu theorem in mathematics states that linking number equals twist plus writhe. Similarly, the helicity of a flux tube equals twist helicity plus writhe helicity \citep{BergerPrior2006}. Therefore, we obtain the twist helicity simply by subtracting the writhe helicity from the total helicity. We can also obtain the mean twist by dividing the twist helicity by flux squared.


In many cases, however, there will be a degree of disorder in the geometry of the field lines, and choosing an axis might not be straightforward. In order to do so for the simulation, we first find on the boundary the flux-weighted centre point of the field lines entering the volume. Then, we calculate the field line from that centre point.  

In our simulation, the regions 2, 4, 7, and 8 have the shape of flux tubes. For these groups, the decomposition of self helicity into twist and writhe is given in table \ref{tabletwistwrithe}.

\begin{table}[!hb]
\begin{center}
\begin{tabular}{||c c c c||} 
 \hline
{\it Group}  & Self Helicity/Flux$^2$ &Twist & Writhe \\ [0.5ex] 
 \hline\hline
 2 &  -0.23 & -0.37 & 0.14 \\
 \hline
 4  & -0.060 & 0.058 & -0.0024 \\
 \hline
 7  & -0.33 & - 0.41 & 0.085 \\
 \hline
 8  & 0.26 & 0.31 & -0.056 \\[0.5ex] 
 \hline
\end{tabular}\label{tabletwistwrithe}
\caption{ Decomposition of self helicity into twist and writhe (divided by flux squared) for the four regions for which an axis can be identified.}
\end{center}
\end{table}

We note two quick remarks on the division of twist and writhe. First, in our simulation, the writhe is always of the opposite sign of the twist. This might be specific to this simulation but it might also represent a general tendency of highly twisted flux ropes to evolve in a way that their total helicity is reduced by the presence and development of writhe. Further investigations with different simulations will be needed to determine the generality of this result. Second, for region 2 (the one with the largest net magnetic flux), neglecting writhe result in an overestimation of the helicity by about 60\%. For other regions, the difference is smaller but because their magnetic flux and total self helicity are small, the overestimation for region 2 is the key result.

\section{Discussions and Conclusions} \label{sec:conclusion}

Magnetic clouds in the solar wind can have complicated magnetic structures. While the focus of many past studies and models has been on twist, writhe is another form of self-helicity and the total magnetic helicity is the sum of the mutual and self helicities. While magnetic helicity is a conserved quantity in a closed system, there is no reason for twist to be conserved as a CME propagates. The two key problems that triggered this investigation are to know i) what is the repartition of magnetic helicity within CMEs between various forms and ii) is it accurate to focus our modeling and fitting efforts on a highly twisted magnetic flux rope.

 In this paper, we have developed a technique to quickly quantify magnetic helicity in a 3-D domain, whether spherical or cubic. First, we suggest dividing up the domain and magnetic field lines into different magnetic regions by using by separatrix surfaces or quasi-separatrix layers. Thus a simple twisted flux tube model may only have one region, while other models may have a more fragmented structure. Next, each magnetic region will have a self helicity and mutual helicities with other regions. The self helicity can be divided into twist and writhe. We suggest calculating the helicity by adding up the winding numbers of pairs of field lines threading the region. The winding numbers can be found using the absolute helicity method \citep{Berger:2018}, which involves finding the poloidal field on the boundary of the region.

We applied this technique to a 3-D simulation of a CME initiated with an out-of-equilibrium highly-twisted flux rope and propagated self-consistently from the Sun to the 1~au with a background IMF. Near 1~au, the magnetic field domain can be divided into eight main regions, with the central magnetic flux rope being the core and the region with the largest magnetic flux. By applying our techniques, we found that i) the self-helicity of this magnetic cloud was only about 60\% of its twist, ii) the total helicity in the domain was only about 70\% of the total self-helicity. Some may argue that, for this one simulation, only taking into consideration the twist would be an acceptable solution (error of about 14\%). However, we point out that this is the case only because the two sources of errors, namely neglecting the writhe and neglecting the mutual helicities, are of similar magnitude and of opposite sign and therefore cancel each other. In fact, the writhe is of comparable magnitude as the twist (being about 38\%) and that the mutual helicity is also of comparable magnitude to the self-helicity (being about 40\%). Neglecting these contributions to the total magnetic helicity is therefore not warranted and studying the evolution of the CME magnetic field without taking into consideration writhe and mutual helicity is prawn to cause errors. 

Fitting techniques at 1~au typically only focus on quantifying the twist of the magnetic field based on {\it in situ} measurements. We can think of three potential scenarios: the techniques do give an appropriate estimate of the CME twist, they give an appropriate estimate of the CME magnetic helicity, or they do not give an appropriate estimate at all. In \citet{AlHaddad:2011} and \citet{AlHaddad:2019}, we showed how the presence of writhe or mutual helicity in CMEs even with very low twist can create {\it in situ} profiles that mimic the characteristics of highly twisted magnetic flux ropes. It is therefore possible that these techniques give an appropriate estimate of the total helicity but attribute all of the helicity to twist, neglecting writhe and mutual helicity. If that is the case, they are appropriate to investigate the evolution of the total helicity from the Sun to the Earth, but not helicity transformation and they give us no information about the exact geometry of the magnetic field within magnetic clouds near 1~au. If, on the opposite, those techniques are able to quantify the twist only of the structure, it may be as much as a factor of 2 off for the total helicity and those techniques may not be appropriate to investigate conservation of magnetic helicity. Obviously, if those techniques do not reproduce either the twist or the total helicity, their usage is very limited. 

We note that the paradigm of highly twisted magnetic flux ropes is so prevalent that visual representations of CMEs often only include twist, whereas it is clear that writhe and mutual helicities are important components. An effort to improve upon these visualizations was recently proposed in \citet{AlHaddad:2024}. The next step for this work is to reproduce these calculations for different simulations, ideally some not initiated with a highly twisted magnetic flux ropes (for example a breakout model) or with more complex versions of the flux ropes \citep[for example the work by][]{Torok:2018}. In parallel, these techniques should be applied at multiple timesteps within the simulation to investigate the transformation of the magnetic helicity fro one form to another and to determine how it depends on various CME and background properties.


\begin{acknowledgments}
The authors thank Florian Regnault and Antoine Strugarek for providing the simulation data necessary to perform the analysis. The authors acknowledge the support of the following grants: NSF-EAGER: AGS2027322, NSF- Solar-Terresterial Core: AGS1954983, NASA-ECIP: 80NSSC21K0463. 
\end{acknowledgments}


\begin{thebibliography}{}
\expandafter\ifx\csname natexlab\endcsname\relax\def\natexlab#1{#1}\fi
\providecommand{\url}[1]{\href{#1}{#1}}
\providecommand{\dodoi}[1]{doi:~\href{http://doi.org/#1}{\nolinkurl{#1}}}
\providecommand{\doeprint}[1]{\href{http://ascl.net/#1}{\nolinkurl{http://ascl.net/#1}}}
\providecommand{\doarXiv}[1]{\href{https://arxiv.org/abs/#1}{\nolinkurl{https://arxiv.org/abs/#1}}}

\bibitem[{{Al-Haddad} \& {Lugaz}(2024)}]{AlHaddad:2024}
{Al-Haddad}, N., \& {Lugaz}, N. 2024, Space Sci. Rev., {{\it in revision}}

\bibitem[{{Al-Haddad} {et~al.}(2019){Al-Haddad}, {Poedts}, {Roussev},
  {Farrugia}, {Yu}, \& {Lugaz}}]{AlHaddad:2019}
{Al-Haddad}, N., {Poedts}, S., {Roussev}, I., {et~al.} 2019, Astrophys. J.,
  870, 100, \dodoi{10.3847/1538-4357/aaf38d}

\bibitem[{{Al-Haddad} {et~al.}(2011){Al-Haddad}, {Roussev}, {M{\"o}stl},
  {Jacobs}, {Lugaz}, {Poedts}, \& {Farrugia}}]{AlHaddad:2011}
{Al-Haddad}, N., {Roussev}, I.~I., {M{\"o}stl}, C., {et~al.} 2011, Astrophys.
  Journ. Lett., 738, L18, \dodoi{10.1088/2041-8205/738/2/L18}

\bibitem[{{Al-Haddad} {et~al.}(2013){Al-Haddad}, {Nieves-Chinchilla},
  {M{\"o}stl}, {Hidalgo}, {Marubashi}, {Savani}, {Roussev}, {Poedts}, \&
  {Farrugia}}]{AlHaddad:2013}
{Al-Haddad}, N., {Nieves-Chinchilla}, T., {M{\"o}stl}, C., {et~al.} 2013, Solar
  Phys., 284, 129, \dodoi{10.1007/s11207-013-0244-5}

\bibitem[{{Antiochos} {et~al.}(1999){Antiochos}, {DeVore}, \&
  {Klimchuk}}]{Antiochos:1999}
{Antiochos}, S.~K., {DeVore}, C.~R., \& {Klimchuk}, J.~A. 1999, Astrophys. J.,
  510, 485, \dodoi{10.1086/306563}

\bibitem[{Berger \& Hornig(2018)}]{Berger:2018}
Berger, M., \& Hornig, G. 2018, Journal of Physics A: Mathematical and
  Theoretical, 51, \dodoi{10.1088/1751-8121/aaea88}

\bibitem[{Berger(2009)}]{Berger:2009}
Berger, M.~A. 2009, Topological Magnetohydrodynamics and Astrophysics (New
  York, NY: Springer New York), 9268--9282,
  \dodoi{10.1007/978-0-387-30440-3_557}

\bibitem[{Berger \& Field(1984)}]{Berger:1984}
Berger, M.~A., \& Field, G.~B. 1984, Journal of Fluid Mechanics, 147, 133

\bibitem[{Berger \& Prior(2006)}]{BergerPrior2006}
Berger, M.~A., \& Prior, C. 2006, J. Phys. A: Math. Gen., 39, 8321

\bibitem[{{Berger, M.A.}(1989)}]{Berger:1989}
{Berger, M.A.} 1989, in Reconnection in Space Plasma, ed. T.~{Guyenne} \&
  J.~{Hunt}

\bibitem[{{Burlaga} {et~al.}(1981){Burlaga}, {Sittler}, {Mariani}, \&
  {Schwenn}}]{Burlaga:1981}
{Burlaga}, L., {Sittler}, E., {Mariani}, F., \& {Schwenn}, R. 1981, J. Geophys.
  Res., 86, 6673

\bibitem[{{Burlaga} \& {King}(1979)}]{Burlaga:1979}
{Burlaga}, L.~F., \& {King}, J.~H. 1979, J. Geophys. Res., 84, 6633,
  \dodoi{10.1029/JA084iA11p06633}

\bibitem[{Campbell \& Berger(2014)}]{Campbell:2014}
Campbell, J., \& Berger, M.~A. 2014, J. Phys.: Conf. Ser., 544, 012001,
  \dodoi{10.1088/1742-6596/544/1/012001}

\bibitem[{{Chen}(2011)}]{Chen:2011}
{Chen}, P.~F. 2011, Living Reviews in Solar Physics, 8, 1,
  \dodoi{10.12942/lrsp-2011-1}

\bibitem[{{Cheng} {et~al.}(2013){Cheng}, {Zhang}, {Ding}, {Liu}, \&
  {Poomvises}}]{Cheng:2013}
{Cheng}, X., {Zhang}, J., {Ding}, M.~D., {Liu}, Y., \& {Poomvises}, W. 2013,
  \apj, 763, 43, \dodoi{10.1088/0004-637X/763/1/43}

\bibitem[{{Dacie} {et~al.}(2018){Dacie}, {T{\"o}r{\"o}k}, {D{\'e}moulin},
  {Linton}, {Downs}, {van Driel-Gesztelyi}, {Long}, \& {Leake}}]{Dacie:2018}
{Dacie}, S., {T{\"o}r{\"o}k}, T., {D{\'e}moulin}, P., {et~al.} 2018, Astrophys.
  Journ. Lett., 862, 117, \dodoi{10.3847/1538-4357/aacce3}

\bibitem[{{Dasso} {et~al.}(2006){Dasso}, {Mandrini}, {D{\'e}moulin}, \&
  {Luoni}}]{Dasso:2006}
{Dasso}, S., {Mandrini}, C.~H., {D{\'e}moulin}, P., \& {Luoni}, M.~L. 2006,
  Astron. Astrophys., 455, 349, \dodoi{10.1051/0004-6361:20064806}

\bibitem[{{D{\'e}moulin} \& {Aulanier}(2010)}]{Demoulin:2010}
{D{\'e}moulin}, P., \& {Aulanier}, G. 2010, Astrophys. J., 718, 1388,
  \dodoi{10.1088/0004-637X/718/2/1388}

\bibitem[{{D{\'e}moulin} \& {Berger}(2003)}]{Demoulin:2003}
{D{\'e}moulin}, P., \& {Berger}, M.~A. 2003, Solar Phys., 215, 203,
  \dodoi{10.1023/A:1025679813955}

\bibitem[{{Fan}(2005)}]{Fan:2005}
{Fan}, Y. 2005, Astrophys. J., 630, 543, \dodoi{10.1086/431733}

\bibitem[{{Farrugia} {et~al.}(2011){Farrugia}, {Berdichevsky}, {M{\"o}stl},
  {Galvin}, {Leitner}, {Popecki}, {Simunac}, {Opitz}, {Lavraud}, {Ogilvie},
  {Veronig}, {Temmer}, {Luhmann}, \& {Sauvaud}}]{Farrugia:2011}
{Farrugia}, C.~J., {Berdichevsky}, D.~B., {M{\"o}stl}, C., {et~al.} 2011, J.
  Atmos. Solar-Terr. Phys., 73, 1254, \dodoi{10.1016/j.jastp.2010.09.011}

\bibitem[{{Forbes} {et~al.}(2006){Forbes}, {Linker}, {Chen}, {Cid}, {K{\'o}ta},
  {Lee}, {Mann}, {Miki{\'c}}, {Potgieter}, {Schmidt}, {Siscoe}, {Vainio},
  {Antiochos}, \& {Riley}}]{Forbes:2006}
{Forbes}, T.~G., {Linker}, J.~A., {Chen}, J., {et~al.} 2006, Space Sci. Rev.,
  123, 251, \dodoi{10.1007/s11214-006-9019-8}

\bibitem[{Gibson \& Fan(2006)}]{Gibson:2006}
Gibson, S., \& Fan, Y. 2006, Journal of Geophysical Research: Space Physics,
  111

\bibitem[{{Gibson}(2018)}]{Gibson:2018}
{Gibson}, S.~E. 2018, Living Reviews in Solar Physics, 15, 7,
  \dodoi{10.1007/s41116-018-0016-2}

\bibitem[{{Green} {et~al.}(2018){Green}, {T{\"o}r{\"o}k}, {Vr{\v{s}}nak},
  {Manchester}, \& {Veronig}}]{Green:2018}
{Green}, L.~M., {T{\"o}r{\"o}k}, T., {Vr{\v{s}}nak}, B., {Manchester}, W., \&
  {Veronig}, A. 2018, \ssr, 214, 46, \dodoi{10.1007/s11214-017-0462-5}

\bibitem[{Hopcroft \& Tarjan(1973)}]{Hopcroft73}
Hopcroft, J., \& Tarjan, R. 1973, Communications of the ACM, 16, 372

\bibitem[{{Hu} {et~al.}(2017){Hu}, {Linton}, {Wood}, {Riley}, \&
  {Nieves-Chinchilla}}]{Hu:2017}
{Hu}, Q., {Linton}, M.~G., {Wood}, B.~E., {Riley}, P., \& {Nieves-Chinchilla},
  T. 2017, Solar Phys., 292, 171, \dodoi{10.1007/s11207-017-1195-z}

\bibitem[{{Hu} \& {Sonnerup}(2002)}]{Hu:2002}
{Hu}, Q., \& {Sonnerup}, B.~U.~{\"O}. 2002, J. Geophys. Res., 107, 1142,
  \dodoi{10.1029/2001JA000293}

\bibitem[{Hu {et~al.}(2022)Hu, Zhu, He, Qiu, Jian, \& Prasad}]{Hu:2022}
Hu, Q., Zhu, C., He, W., {et~al.} 2022, The Astrophysical Journal, 934, 50,
  \dodoi{10.3847/1538-4357/ac7803}

\bibitem[{{Jin} {et~al.}(2017){Jin}, {Manchester}, {van der Holst}, {Sokolov},
  {T{\'o}th}, {Mullinix}, {Taktakishvili}, {Chulaki}, \& {Gombosi}}]{Jin:2017}
{Jin}, M., {Manchester}, W.~B., {van der Holst}, B., {et~al.} 2017, \apj, 834,
  173, \dodoi{10.3847/1538-4357/834/2/173}

\bibitem[{{Kilpua} {et~al.}(2011){Kilpua}, {Jian}, {Li}, {Luhmann}, \&
  {Russell}}]{Kilpua:2011}
{Kilpua}, E.~K.~J., {Jian}, L.~K., {Li}, Y., {Luhmann}, J.~G., \& {Russell},
  C.~T. 2011, Journal of Atmospheric and Solar-Terrestrial Physics, 73, 1228,
  \dodoi{10.1016/j.jastp.2010.10.012}

\bibitem[{Klenin \& Langowski(2000)}]{Klenin:2000}
Klenin, K., \& Langowski, J. 2000, Biopolymers: Original Research on
  Biomolecules, 54, 307

\bibitem[{{Kusano} {et~al.}(2004){Kusano}, {Maeshiro}, {Yokoyama}, \&
  {Sakurai}}]{Kusano:2004}
{Kusano}, K., {Maeshiro}, T., {Yokoyama}, T., \& {Sakurai}, T. 2004, \apj, 610,
  537, \dodoi{10.1086/421547}

\bibitem[{Ledvina {et~al.}(2023)Ledvina, Palmerio, Kay, Al-Haddad, \&
  Riley}]{Ledvina:2023}
Ledvina, V., Palmerio, E., Kay, C., Al-Haddad, N., \& Riley, P. 2023, Astronomy
  Astrophysics, 673, A96, \dodoi{10.1051/0004-6361/202245445}

\bibitem[{{Lemen} {et~al.}(2012){Lemen}, {Title}, {Akin}, {Boerner}, {Chou},
  {Drake}, {Duncan}, {Edwards}, {Friedlaender}, {Heyman}, {Hurlburt}, {Katz},
  {Kushner}, {Levay}, {Lindgren}, {Mathur}, {McFeaters}, {Mitchell}, {Rehse},
  {Schrijver}, {Springer}, {Stern}, {Tarbell}, {Wuelser}, {Wolfson}, {Yanari},
  {Bookbinder}, {Cheimets}, {Caldwell}, {Deluca}, {Gates}, {Golub}, {Park},
  {Podgorski}, {Bush}, {Scherrer}, {Gummin}, {Smith}, {Auker}, {Jerram},
  {Pool}, {Soufli}, {Windt}, {Beardsley}, {Clapp}, {Lang}, \&
  {Waltham}}]{Lemen:2012}
{Lemen}, J.~R., {Title}, A.~M., {Akin}, D.~J., {et~al.} 2012, Solar Phys., 275,
  17, \dodoi{10.1007/s11207-011-9776-8}

\bibitem[{{Lepping} {et~al.}(1990){Lepping}, {Burlaga}, \&
  {Jones}}]{Lepping:1990}
{Lepping}, R.~P., {Burlaga}, L.~F., \& {Jones}, J.~A. 1990, J. Geophys. Res.,
  95, 11957

\bibitem[{{Lopez}(1987)}]{Lopez:1987}
{Lopez}, R.~E. 1987, J. Geophys. Res., 92, 11189,
  \dodoi{10.1029/JA092iA10p11189}

\bibitem[{{Lugaz} {et~al.}(2018){Lugaz}, {Farrugia}, {Winslow}, {Al-Haddad},
  {Galvin}, {Nieves-Chinchilla}, {Lee}, \& {Janvier}}]{Lugaz:2018}
{Lugaz}, N., {Farrugia}, C.~J., {Winslow}, R.~M., {et~al.} 2018, \apjl, 864,
  L7, \dodoi{10.3847/2041-8213/aad9f4}

\bibitem[{{Lugaz} {et~al.}(2017){Lugaz}, {Temmer}, {Wang}, \&
  {Farrugia}}]{Lugaz:2017}
{Lugaz}, N., {Temmer}, M., {Wang}, Y., \& {Farrugia}, C.~J. 2017, Solar Phys.,
  292, 64, \dodoi{10.1007/s11207-017-1091-6}

\bibitem[{{Lugaz} {et~al.}(2024){Lugaz}, {Zhuang}, {Scolini}, {Al-Haddad},
  {Farrugia}, {Winslow}, {Regnault}, {M{\"o}stl}, {Davies}, \&
  {Galvin}}]{Lugaz:2024}
{Lugaz}, N., {Zhuang}, B., {Scolini}, C., {et~al.} 2024, Astrophys. J., 962,
  193, \dodoi{10.3847/1538-4357/ad17b9}

\bibitem[{{Lynch} {et~al.}(2008){Lynch}, {Antiochos}, {DeVore}, {Luhmann}, \&
  {Zurbuchen}}]{Lynch:2008}
{Lynch}, B.~J., {Antiochos}, S.~K., {DeVore}, C.~R., {Luhmann}, J.~G., \&
  {Zurbuchen}, T.~H. 2008, Astrophys. J., 683, 1192, \dodoi{10.1086/589738}

\bibitem[{MacTaggart \& Prior(2021)}]{Mactaggart:2021}
MacTaggart, D., \& Prior, C. 2021, Geophysical \& Astrophysical Fluid Dynamics,
  115, 85

\bibitem[{{Manchester} {et~al.}(2017){Manchester}, {Kilpua}, {Liu}, {Lugaz},
  {Riley}, {T{\"o}r{\"o}k}, \& {Vr{\v s}nak}}]{Manchester:2017}
{Manchester}, W., {Kilpua}, E.~K.~J., {Liu}, Y.~D., {et~al.} 2017, Space Sci.
  Rev., 212, 1159, \dodoi{10.1007/s11214-017-0394-0}

\bibitem[{{Marubashi}(1986)}]{Marubashi:1986}
{Marubashi}, K. 1986, Adv. Space Res., 6, 335,
  \dodoi{10.1016/0273-1177(86)90172-9}

\bibitem[{Moffatt(1969)}]{Moffatt69}
Moffatt, H.~K. 1969, Journal of Fluid Mechanics, 35, 117

\bibitem[{{M{\"o}stl} {et~al.}(2008){M{\"o}stl}, {Miklenic}, {Farrugia},
  {Temmer}, {Veronig}, {Galvin}, {Vr{\v s}nak}, \& {Biernat}}]{Moestl:2008}
{M{\"o}stl}, C., {Miklenic}, C., {Farrugia}, C.~J., {et~al.} 2008, Annales
  Geophysicae, 26, 3139, \dodoi{10.5194/angeo-26-3139-2008}

\bibitem[{{Nieves-Chinchilla} {et~al.}(2018){Nieves-Chinchilla}, {Linton},
  {Hidalgo}, \& {Vourlidas}}]{Nieves:2018b}
{Nieves-Chinchilla}, T., {Linton}, M.~G., {Hidalgo}, M.~A., \& {Vourlidas}, A.
  2018, \apj, 861, 139, \dodoi{10.3847/1538-4357/aac951}

\bibitem[{{Nieves-Chinchilla} {et~al.}(2016){Nieves-Chinchilla}, {Linton},
  {Hidalgo}, {Vourlidas}, {Savani}, {Szabo}, {Farrugia}, \& {Yu}}]{Nieves:2016}
{Nieves-Chinchilla}, T., {Linton}, M.~G., {Hidalgo}, M.~A., {et~al.} 2016,
  \apj, 823, 27, \dodoi{10.3847/0004-637X/823/1/27}

\bibitem[{{Nindos} {et~al.}(2003){Nindos}, {Zhang}, \& {Zhang}}]{Nindos:2003}
{Nindos}, A., {Zhang}, J., \& {Zhang}, H. 2003, \apj, 594, 1033,
  \dodoi{10.1086/377126}

\bibitem[{{Pariat} {et~al.}(2017){Pariat}, {Leake}, {Valori}, {Linton},
  {Zuccarello}, \& {Dalmasse}}]{Pariat:2017}
{Pariat}, E., {Leake}, J.~E., {Valori}, G., {et~al.} 2017, Astron. Astrophys.,
  601, A125, \dodoi{10.1051/0004-6361/201630043}

\bibitem[{{Patsourakos} {et~al.}(2013){Patsourakos}, {Vourlidas}, \&
  {Stenborg}}]{Patsourakos:2013}
{Patsourakos}, S., {Vourlidas}, A., \& {Stenborg}, G. 2013, \apj, 764, 125,
  \dodoi{10.1088/0004-637X/764/2/125}

\bibitem[{{Patsourakos} {et~al.}(2020){Patsourakos}, {Vourlidas},
  {T{\"o}r{\"o}k}, {Kliem}, {Antiochos}, {Archontis}, {Aulanier}, {Cheng},
  {Chintzoglou}, {Georgoulis}, {Green}, {Leake}, {Moore}, {Nindos}, {Syntelis},
  {Yardley}, {Yurchyshyn}, \& {Zhang}}]{Patsourakos:2020}
{Patsourakos}, S., {Vourlidas}, A., {T{\"o}r{\"o}k}, T., {et~al.} 2020, Space
  Sci. Rev., 216, 131, \dodoi{10.1007/s11214-020-00757-9}

\bibitem[{Pevtsov {et~al.}(2014)Pevtsov, Berger, Canfield, A., Norton, \& van
  Driel-Gesztelyi}]{Pevtsov2014}
Pevtsov, A.~A., Berger, M.~A., Canfield, R.~C., {et~al.} 2014, Space science
  Reviews, 186, 285, \dodoi{10.1007/s11214-014-0082-2}

\bibitem[{Priest \& D\'emoulin(1995)}]{Priest:1995}
Priest, E., \& D\'emoulin, P. 1995, JGR Space Physics, 100, A12,
  \dodoi{10.1029/95JA02740}

\bibitem[{Prior {et~al.}(2020)Prior, Hawkes, \& Berger}]{Prior:2020}
Prior, C., Hawkes, G., \& Berger, M. 2020, Astronomy \& Astrophysics, 635,
  \dodoi{10.1051/0004-6361/201936675}

\bibitem[{Prior \& Yeates(2014)}]{prioryeates14}
Prior, C., \& Yeates, A. 2014, Astrophys. J., 787, 100

\bibitem[{Prior \& Yeates(2021)}]{Prior:2021}
---. 2021, Journal of Physics A: Mathematical and Theoretical, 54, 465701

\bibitem[{{Regnault} {et~al.}(2023{\natexlab{a}}){Regnault}, {Al-Haddad},
  {Lugaz}, {Farrugia}, {Yu}, {Davies}, {Galvin}, \& {Zhuang}}]{Regnault:2023b}
{Regnault}, F., {Al-Haddad}, N., {Lugaz}, N., {et~al.} 2023{\natexlab{a}},
  Astrophys. J., 957, 49, \dodoi{10.3847/1538-4357/acef16}

\bibitem[{{Regnault} {et~al.}(2024){Regnault}, {Al-Haddad}, {Lugaz},
  {Farrugia}, {Zhuang}, {Yu}, \& {Strugarek}}]{Regnault:2024b}
---. 2024, Astrophys. Journ. Lett., 966, L17, \dodoi{10.3847/2041-8213/ad3806}

\bibitem[{{Regnault} {et~al.}(2023{\natexlab{b}}){Regnault}, {Strugarek},
  {Janvier}, {Auch{\`e}re}, {Lugaz}, \& {Al-Haddad}}]{Regnault:2023a}
{Regnault}, F., {Strugarek}, A., {Janvier}, M., {et~al.} 2023{\natexlab{b}},
  Astron. Astrophys., 670, A14, \dodoi{10.1051/0004-6361/202244483}

\bibitem[{{Riley} {et~al.}(2001){Riley}, {Linker}, \& {Miki{\'c}}}]{Riley:2001}
{Riley}, P., {Linker}, J.~A., \& {Miki{\'c}}, Z. 2001, J. Geophys. Res., 106,
  15889, \dodoi{10.1029/2000JA000121}

\bibitem[{{Romashets} \& {Vandas}(2003)}]{Romashets:2003}
{Romashets}, E.~P., \& {Vandas}, M. 2003, Geophys. Res. Lett., 30, 200000,
  \dodoi{10.1029/2003GL017692}

\bibitem[{{Roussev} {et~al.}(2007){Roussev}, {Lugaz}, \&
  {Sokolov}}]{Roussev:2007}
{Roussev}, I.~I., {Lugaz}, N., \& {Sokolov}, I.~V. 2007, Astrophys. Journ.
  Lett., 668, L87, \dodoi{10.1086/522588}

\bibitem[{Tassev \& Savcheva(2017)}]{Tassev:2017}
Tassev, S., \& Savcheva, A. 2017, The Astrophysical Journal, 840, 89,
  \dodoi{10.3847/1538-4357/aa6f06}

\bibitem[{Temmer(2023)}]{Temmer:2023}
Temmer, M.~e. 2023, Adv. Space Res., \dodoi{10.1016/j.asr.2023.07.003}

\bibitem[{{Titov}(2007)}]{Titov:2007}
{Titov}, V.~S. 2007, Astrophys. J., 660, 863, \dodoi{10.1086/512671}

\bibitem[{{Titov} \& {D{\'e}moulin}(1999)}]{Titov:1999}
{Titov}, V.~S., \& {D{\'e}moulin}, P. 1999, Astron. Astrophys., 351, 707

\bibitem[{{Titov} {et~al.}(2014){Titov}, {T{\"o}r{\"o}k}, {Mikic}, \&
  {Linker}}]{Titov:2014}
{Titov}, V.~S., {T{\"o}r{\"o}k}, T., {Mikic}, Z., \& {Linker}, J.~A. 2014,
  Astrophys. J., 790, 163, \dodoi{10.1088/0004-637X/790/2/163}

\bibitem[{{Torok} {et~al.}(2022){Torok}, {Ben-Nun}, {Downs}, {Titov}, {Caplan},
  \& {Lionello}}]{Torok:2022}
{Torok}, T., {Ben-Nun}, M., {Downs}, C., {et~al.} 2022, in EGU General Assembly
  Conference Abstracts, EGU General Assembly Conference Abstracts, EGU22--2040,
  \dodoi{10.5194/egusphere-egu22-2040}

\bibitem[{{T{\"o}r{\"o}k} \& {Kliem}(2003)}]{Torok:2003}
{T{\"o}r{\"o}k}, T., \& {Kliem}, B. 2003, Astron. Astrophys., 406, 1043,
  \dodoi{10.1051/0004-6361:20030692}

\bibitem[{T{\"o}r{\"o}k {et~al.}(2018)T{\"o}r{\"o}k, Downs, Linker, Lionello,
  Titov, Miki{\'c}, Riley, Caplan, \& Wijaya}]{Torok:2018}
T{\"o}r{\"o}k, T., Downs, C., Linker, J.~A., {et~al.} 2018, The Astrophysical
  Journal, 856, 75

\bibitem[{{Vourlidas} {et~al.}(2000){Vourlidas}, {Subramanian}, {Dere}, \&
  {Howard}}]{Vourlidas:2000}
{Vourlidas}, A., {Subramanian}, P., {Dere}, K.~P., \& {Howard}, R.~A. 2000,
  Astrophys. J., 534, 456

\bibitem[{Xiao {et~al.}(2023)Xiao, Prior, \& Yeates}]{Xiao:2023}
Xiao, D., Prior, C.~B., \& Yeates, A.~R. 2023, Journal of Physics A:
  Mathematical and Theoretical, 56, 205201

\bibitem[{{Yashiro} {et~al.}(2004){Yashiro}, {Gopalswamy}, {Michalek},
  {St.~Cyr}, {Plunkett}, {Rich}, \& {Howard}}]{Yashiro:2004}
{Yashiro}, S., {Gopalswamy}, N., {Michalek}, G., {et~al.} 2004, Journal of
  Geophysical Research (Space Physics), 14, \dodoi{10.1029/2003JA010282}

\bibitem[{{Zhuang} {et~al.}(2022){Zhuang}, {Lugaz}, {Temmer}, {Gou}, \&
  {Al-Haddad}}]{Zhuang:2022}
{Zhuang}, B., {Lugaz}, N., {Temmer}, M., {Gou}, T., \& {Al-Haddad}, N. 2022,
  Astrophys. J., 933, 169, \dodoi{10.3847/1538-4357/ac75d4}

\end{thebibliography}
\end{document}